\documentclass[a4paper,11pt]{article}
\pdfoutput=1
\usepackage{jheppub}
\usepackage{natbib}
\usepackage[T1]{fontenc} 
\usepackage{fancyhdr}
\usepackage{amsthm}
\usepackage{enumerate}
\usepackage{subfig,color}
\usepackage[export]{adjustbox}
\newcommand{\beq}{\begin{equation}}
\newcommand{\eeq}{\end{equation}}
\newcommand{\bea}{\begin{eqnarray}}
\newcommand{\eea}{\end{eqnarray}}

\title{Cosmological Tests of Everpresent $\Lambda$}

\author[a,b]{Nosiphiwo Zwane}
\author[a,b]{, Niayesh Afshordi}
\author[a,b,c]{ and Rafael D. Sorkin}

\affiliation[a]{Perimeter Institute for Theoretical Physics, 31 Caroline St. N., Waterloo, ON, N2L 2Y5, Canada}
\affiliation[b]{Department of Physics and Astronomy, University of Waterloo, Waterloo, ON, N2L 3G1, Canada}
\affiliation[c]{Department of Physics, Syracuse University, Syracuse, NY 13244-1130, U.S.A.}

\abstract{
Everpresent $\Lambda$ is a cosmological scenario in which
the observed cosmological ``constant'' $\Lambda$ fluctuates between positive and negative values with a vanishing mean, and with a magnitude comparable to the critical density at any epoch. In accord with a longstanding heuristic prediction of {\it causal set} theory, it postulates that $\Lambda$ is a {\it stochastic} function of cosmic time that will vary from one realization of the scenario to another. Herein, we consider two models of ``dark energy'' that exhibit these features.
Via Monte Carlo Markov chains, we explore the space of cosmological parameters {\it\/and\/} the set of stochastic realizations of these models, finding that Everpresent $\Lambda$ can fit current cosmological observations as well as the $\Lambda$CDM model does.
Furthermore, it removes the observational tensions that $\Lambda$CDM experiences in relation to low redshift measurements of the Hubble constant, and to the Baryonic Acoustic Oscillations (BAO) in Lyman-$\alpha$ forest at $z\sim 2-3$. It does not, however, help significantly
with the early growth of ultramassive black holes, or with the Lithium problem in Big Bang Nucleosynthesis.
Future measurements of ``dark energy'' at high redshifts will further test the viability of Everpresent $\Lambda$ as an alternative to the $\Lambda$CDM cosmology.}

\begin{document}
\date{\today}
\maketitle
\flushbottom

\section{Introduction}
The causal set programme for quantum gravity gave rise to the prediction that fluctuations in the cosmological ``constant'' must be present at all epochs, with an amplitude inversely proportional to the square root of spacetime volume \cite{everP0a,everP0b,everP0c,everP0d,everP0e,everP1,everP2,everP3,Ng}. The fluctuations are a quantum mechanical result of spacetime being fundamentally discrete, which entails uncertainties in volume.  The cosmological ``constant'' $\Lambda$ is consequently a function of time (and in principle, space), hence the Einstein's field equations for General Relativity do not hold. In this paper, we examine some cosmological tests of one possible model of these fluctuations, and demonstrate that, contrary to the widespread belief that ``dark energy'' must be insignificant at high redshifts, the model is consistent with current observational data.


In Section \ref{sec1}, we develop a stochastic phenomenological model of Everpresent $\Lambda$, or dark energy, which satisfies the broad expectations from causal sets and quantum gravity.  Moreover, the models (there are actually two of them) admit a fully covariant and causal perturbation theory, allowing us to consistently derive predictions for the anisotropies in the cosmic microwave background (CMB).
%



In Section \ref{sec4}, we fit the model to the CMB
using Planck 2015 data \cite{2016A&A...594A..13P}. Even though Everpresent $\Lambda$ is a stochastic model, a Monte Carlo exploration of the space of likely dark-energy histories using the $\mathtt{CAMB}$ code finds several histories that are excellent fits to the Planck 2015 data. As could be expected with a stochastic model, the data does not single out just one good history of dark energy, but it selects a range of them, basically the histories for which $\Omega_{de} \simeq 0.7$ at $z\simeq 0$ and $\Omega_{de} \simeq 0$ at $z \simeq 1000$, at 5\% level.
%
%
%
Since Everpresent $\Lambda$ is a stochastic model, we pay special attention to how one should compare it to conventional deterministic dark-energy models, and to what extent good fits are (or will be) contrived.

In Section \ref{local_hubble}, we examine the recent tension between the direct measurements of the local Hubble constant from the cosmological distance ladder on one hand, and the best fit values from CMB \cite{LocalHub,2017NatAs...1E.169F} on the other hand. We shall see that this tension is removed in the Everpresent $\Lambda$ model, thanks to its stochastic nature.


Section \ref{sec3} compares Everpresent $\Lambda$ predictions for volume weighted distance $D_v$ with Baryon acoustic oscillation (BAO) data from several surveys. BAO observations provide an independent measure of the expansion of the universe which complements supernovae data \cite{supeV} in tests of dark energy models. In particular, results from the Baryon Oscillation Spectroscopic Survey (BOSS) \cite{BAO1,BAO2} suggest that $\Omega_{de}(z)$ is negative at $z=2.34$ in $\sim 2.5 \sigma$ tension with standard $\Lambda$CDM. A number of other models that Aubourg \textit{et al.} \cite{BAO3} examined fail to fit the 2013 BAO data unless one assumes that
$\Omega_{de}<0$ at $z\sim 2-3$~. This is not a problem for
Everpresent $\Lambda$, since it allows for $\Omega_{de}(z)$ to change sign roughly once per Hubble time.

In Section \ref{sec5}, we ask whether Everpresent $\Lambda$ can resolve the difficulty that $\Lambda$CDM is thought to have with the occurrence of ultramassive black holes at high redshift. Quasar black holes at $z\sim$ 6 to 7 are more massive than would be expected on the basis of a $\Lambda$CDM cosmology combined with standard or plausible astrophysical assumptions, including a sub-Eddington accretion rate (e.g. \cite{Tpro}) and maximal angular velocity of the black holes.
%
%
Exotic astrophysical processes might increase the accretion rate, allowing the quasars to accumulate mass faster, but such processes have not been observed, leading one to look for cosmological explanations. We find that the stochastic dark-energy histories typical for Everpresent $\Lambda$ can lengthen the age of the cosmos and thereby stretch the available accretion time, but not enough to reach observed masses for maximally spinning (extremal) black holes.


In Section \ref{sec6}, we discuss Everpresent $\Lambda$ as a cosmological solution to the primordial Lithium-7 problem. Big Bang Nucleosynthesis (BBN), predicts the abundance of light cosmological elements (D,$^3$He,$^4$He,$^7$Li) that are produced within the first $20$ minutes after the Big Bang. The abundances of primordial Deuterium and Helium-4 predicted by BBN are in agreement with astronomically observed abundances.  However BBN in a $\Lambda$CDM cosmology predicts about three times more Lithium-7 than is observed. We show that Everpresent $\Lambda$ can ease this tension, but cannot completely
remove it if we take current astrophysical measurement errors at face value.

Finally, Section \ref{sec_conclude} concludes the paper and discusses future prospects for tests of Everpresent $\Lambda$.

But, let us first briefly review the Everpresent $\Lambda$ idea itself.

\section{Everpresent $\Lambda$} \label{sec1}

That the cosmological ``constant'' should vary stochastically was an early heuristic prediction of causal set theory \cite{everP0a,everP0b,everP0c,everP0d,everP0e}.
A concrete model of such an {\it\/Everpresent $\Lambda$} was first proposed by Ahmed \textit{et al.} \cite{everP1}.
Causal set theory \cite{Cset}
assumes spacetime to be discrete at Planck scale, with its elements forming a partially ordered set (ordered by their causal relation) called a causal set.  Roughly speaking, the number of elements determines the spacetime volume, and the order gives rise to the causal structure of the spacetime. The order and number of the elements together give rise to the geometry of spacetime: ``Order + Number = Geometry''!

The equality between number and volume is not exact if Lorentz Invariance holds, but is subject to Poisson fluctuations. In Planck units ($8\pi G=\hbar=c=1$)
\begin{align}
  N \sim V \pm \sqrt{V} . \label{here1}
\end{align}
The term that involves the cosmological constant in the Einstein-Hilbert action is $- \int \Lambda dV$, suggesting that $V$ and $\Lambda$ are canonical conjugates \citep{everP1}. Therefore, based on the Heisenberg uncertainty principle, the quantum fluctuations would obey
\begin{align}
  \Delta \Lambda  \times \Delta V \sim 1 . \label{QM1}
\end{align}
Holding the number $N$ of causet elements fixed does not imply that the volume $V$ is fixed, rather it implies via equations (\ref{here1}) and (\ref{QM1}) that
\begin{align}
  \Delta \Lambda \sim V^{-1/2} . \label{FluC1}
\end{align}
Assuming that $<\Lambda>=0$ (thanks to a yet-unknown solution to the old cosmological constant problem), and taking volume to be roughly the fourth power of the Hubble radius, $H^{-1}$, we obtain
\begin{align}
  \Lambda \sim H^2 = \frac{1}{3} \rho_{c} \ ,
\end{align}
where $\rho_c$ is the critical energy density of the universe. In sum, causal set theory yields a heuristic prediction that the cosmological ``constant'' is of order $H^2$,  that it is consequently ``everpresent'', and that it fluctuates due to Poisson statistics of spacetime causal elements. Here, we will consider two models of such an Everpresent $\Lambda$.

\subsection{Model 1}
Proposed by Ahmed \textit{et al.} \cite{everP1}, this model assumes that the cosmological constant fluctuates due to Poisson fluctuations as discussed above. The universe is assumed to have volume $V(t)$ at some time $t$ (more precisely the volume of the past lightcone of a point), one uses the Friedmann equation(\ref{Frid01})
\begin{align}
  \left(\frac{\dot{a}}{a}\right)^2 = \frac{1}{3} \rho - \frac{\Lambda(t)}{3} \label{Frid01}
\end{align}
to evolve the scale factor to the later time, $a(t+\Delta t)$, and one then updates the 4-volume of the past lightcone using:
\begin{align}
  V(t) &=  \frac{4 \pi}{3} \int_0^t dt' \,a(t')^3 \left( \int_0^{t'} dt'' \frac{1}{a(t'')}  \right)^3, \\
  \ N(t) &= V(t)/\ell_p^4,\\
  \rho_\Lambda (t+1) &= \frac{S(t) + \beta \xi \sqrt{ N(t+1) - N(t)}}{V(t)}.
\end{align}
%
Here, $N_i$ is the number of causet elements at the $i^{th}$ iteration, $\xi$ is a gaussian random number which drives the random walk, and $S$ is the action of ``free spacetime'', $\rho_\Lambda$ being then the action per unit volume.
The parameter $\beta$ controls the magnitude of the fluctuations. When it is too large, the fluctuations in $\rho_{\Lambda}$ are large enough to drive the total effective energy density below zero, and the simulation almost always stops before the universe reaches its present size.
On the other hand, a value of $\beta$ which is too small results in a universe which is too young compared to the oldest known galaxies, which are about $12$ Gyr. Ahmed \textit{et al.} \cite{everP1} found that one needs roughly  $0.01 < \beta < 0.02$ to generate a universe consistent with astrophysical observations. Figure \ref{Reco} shows $\Omega_{de}$ for this model.

\subsection{Model 2}
%
In this model, the cosmological constant is again assumed to be a random function of cosmic time,
but in contrast to Model 1, two phenomenological parameters are introduced to characterize its fluctuations, a parameter $\alpha$ that (as in Model 1) controls the magnitude of its fluctuations, and a second, independent parameter $\mu$ that controls their coherence time (presumed to be comparable to the Hubble time, just as one sees in Model 1).
We conveniently build in the ``everpresent'' nature of $\Lambda$ by expressing the fluctuations directly in terms of  $\Omega_\Lambda$ rather than in terms of the more basic $\rho_\Lambda$ as in Model 1.

Specifically we make the ansatz that
\begin{align}
  \left< \hat{\Omega}_{de}(\lambda_1) \hat{\Omega}_{de}(\lambda_2) \right>
   &= \alpha^2 e^{-\frac{\mu (\lambda_1 -\lambda_2)^2}{2} }
  \label{m2}
\end{align}
where $\lambda=\log(a)$ ($a$ being the cosmological scale factor), and the random variable $\hat{\Omega}_{de}$ is that of the auxiliary classical gaussian random process with mean zero defined by (\ref{m2}). (Since $\hat{\Omega}_{de}$ is gaussian, it is determined fully as a stochastic process by its one- and two-point correlation functions.)
We assume further a spatially flat, homogeneous cosmology and a positive matter density at all times. Since this requires $\Omega_{de}(\lambda)<1$ for all $\lambda$, we cannot directly identify $\hat{\Omega}_{de}$ as being $\Omega_{de}$, but instead we ``compress'' it to the range $(-1,1)$ via the simple symmetric ansatz,
\beq
   \Omega_{de}(\lambda) = \tanh\left[\hat{\Omega}_{de}(\lambda)\right].
\eeq

In summary, $\Omega_{de}$ is the hyperbolic tangent of a random gaussian function of logarithmic time, with zero mean, variance $\alpha$, and a correlation time of ${\rm Hubble~Time}/\sqrt{\mu}$~.

\begin{figure}
\centering
\includegraphics[width=\textwidth]{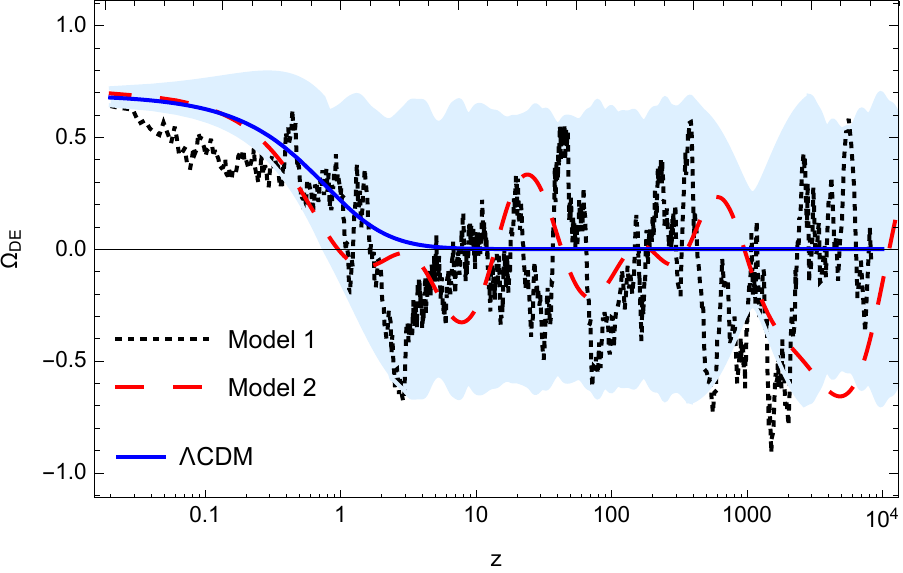}
\caption{Dark Energy history, $\Omega_{de}(z)$, for two Everpresent $\Lambda$ models that fit Planck+BAO data, compared with $\Lambda$CDM model. The shaded area shows the 68\% region of realizations of Mode 2 that fit the data well.} \label{Reco}
\end{figure}

To simulate this model, we Fourier Transform (\ref{m2}) to get
\begin{align}
  \left< \tilde{\Omega}_{de}(\omega_1) \tilde{\Omega}_{de}(\omega_2) \right>
    &= \alpha^2 e^{- \frac{\omega^2}{4\mu}} \sqrt{\frac{\pi}{\mu}}
  \delta(\omega_1 - \omega_2) \ .
\end{align}
We then sample (for a set of frequencies)
$\tilde{\Omega}(\omega)_{de}$'s as independent gaussian variables, and Fourier transform back to $\hat{\Omega}(\lambda)_{de}$, getting a random history of $\Omega_{de}(\lambda)= \tanh \hat{\Omega}(\lambda)_{de}$ as shown in Figure \ref{Reco}.

Because $\hat{\Omega}_{de}$ is a random gaussian function, the reduced $\chi_{red, {\rm model}}^2$ for this model is given by
\begin{align}
  \chi_{red, {\rm model}}^{2}
   \ &= \ \frac{\chi^2_{\rm model}}{\sum_{\omega} \# \omega}
   \ \sim \ 1 \pm \sqrt{\frac{2}{\# \omega}}
\end{align}
where \# is the number of sampled $\omega$'s and,
\begin{align}
  \chi_{\rm model}^2
   &= \sum_{\omega} \frac{\left| \tilde{\Omega}_{de}(\omega) \right| ^2}
   {\left< \left| \tilde{\Omega}_{de}(\omega) \right|^2 \right>} \ .
  \label{chi2e}
\end{align}
We shall use $\chi_{red,{\rm model}}^2$
as a measure of how typical the histories that fit observations may be in the context of Everpresent $\Lambda$. Quantified in this manner, typicality is objective. It thus offers a test which can be applied to a stochastic model such as ours, and which
in principle, could turn out to falsify it.

Since Model 2 is somewhat easier to simulate than Model 1, we shall adopt it in our Monte Carlo Markov Chain (MCMC) exploration of the parameter space below.


\subsection{Inhomogeneities}

We live in an inhomogeneous universe.  Accordingly, any theory of time-dependent dark energy that can be compared with the full set of cosmological observation  must include a consistent model of inhomogeneities, at least at the perturbative level.

Based on their expectations about how such a model would look, Barrow \cite{notgood1} and Zuntz \cite{notgood2} argued for a bound of $10^{-6}$ on the $\Lambda$ fluctuations, which would make them too small to account for the cosmic acceleration today.  They associated a value $\Lambda(x)$ with each spacetime point $x$, and assumed that it would be determined solely by the Poisson fluctuations within the past light cone of $x$, with the implication that the correlation between $\Lambda(a)$ and $\Lambda(b)$ would vanish when the past light cones of $a$ and $b$ were disjoint.  Consequently, separate patches in the CMB sky would reflect uncorrelated $\Omega_\Lambda$ fluctuations, which in turn would limit their amplitude to $\lesssim{}10^{-6}$, from CMB observations.

However, they did not provide any consistent covariant model of the fluctuations in which their plausibility arguments could be verified.\footnote{The fact that a consistent extension of the Everpresent $\Lambda$ idea to include inhomogeneities {\it\/might\/} lead to a scenario like those of \cite{notgood1,notgood2} was recognized from the beginning (see last paragraph of Section IV in \cite{everP1}).  What has been lacking is a concrete (not to mention quantum mechanical) model in which to study this issue.} Moreover, they seem to have had in mind a local stochastic and purely classical model of the fluctuations, whereas Everpresent $\Lambda$ is supposed to be a quantum effect, which necessarily would inherit some of the nonlocality manifested, for example, in violations of the Bell inequalities \cite{Bell:1964kc}.



Here, we shall not attempt to produce a quantum model of the fluctuations, worthy though the project may be.  Rather, we will approach the question of inhomogeneities from a pragmatic standpoint and analyze it in terms of a model which is at least self consistent and covariant.  Basically we shall simply adopt the model provided by the $\mathtt{CAMB}$ software package, which assumes that dark energy is a perfect fluid with perturbations that propagate with the speed of light, as is consistent with the Lorentz symmetry of the underlying causal set. In the cosmological context, this is equivalent to a {\it\/quintessence\/} dark energy \cite{1999PhRvL..82..896Z}, which can be modelled by a scalar field with a canonical kinetic term (but with a potential term that will vary from realization to realization).\footnote{Notice that in Model 2, any difficulties that might be associated with a situation where the pressure falls below minus the energy-density are absent, because the compression of $\Omega_Lambda$ built into Model 2 rules out any such ``phantom'' behavior.} Then, fixing the history of the {\it\/homogeneous\/} dark energy (obtained from either Model 1 or 2 above), and assuming also energy-momentum conservation and adiabatic initial conditions, uniquely determines the behavior of linear perturbations.

This is clearly not a first principles derivation. However, it is a phenomenologically consistent and viable model that captures two main features expected from causal sets: Everpresent temporal fluctuations of $\Lambda$, and local Lorentz symmetry for the inhomogeneities.

\section{CMB and Everpresent $\Lambda$} \label{sec4}
Due to a combination of the quality of the data and the validity of linear perturbation theory, the Cosmic Microwave Background (CMB) anisotropies currently provide the most precise tests of cosmological models. To fit our Everpresent $\Lambda$ model, we use $\mathtt{CosmoMc}$ \cite{ccosmomc} together with $\mathtt{CAMB}$ \cite{ccamb}, some parts of which we had to rewrite in order to accommodate a stochastic dark energy.

As our model is stochastic, even the same values of $\alpha$ and $\mu$ could lead to vastly different geometries of the universe, resulting in e.g., different ages or present-day densities of dark energy. One way to parameterize this variation is through the initial seed-number used to generate the pseudorandom numbers that go into $\Omega_{de}(z)$. Therefore, we shall run CosmoMC with $(\alpha,\mu, seed)$ as additional parameters to be explored by the Monte-Carlo Markov Chain (MCMC). Notice that, unlike with the other cosmological parameters, the likelihood surface will not be a smooth function of $seed$. Nevertheless, an MCMC random walk in parameter-space will randomly sample various dark energy histories and keep the ones that fit the data well. While, due to the very irregular dependence of the likelihood on the $seed$, there is no guarantee that this chain will converge quickly (or at all), we see below that it does find fits to the data comparable to that of standard $\Lambda$CDM cosmology.

\begin{figure}
\centering
\includegraphics[scale=1.3]{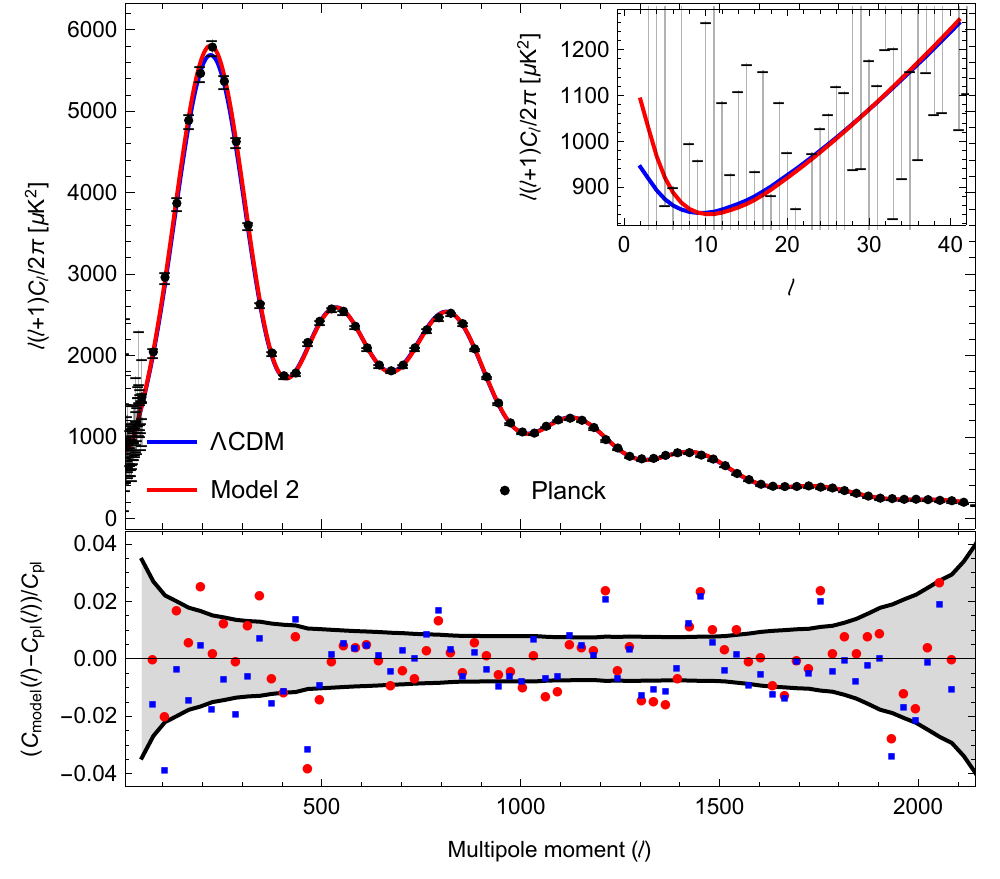}
\caption{Temperature fluctuations for $\Lambda$CDM model and Everpresent $\Lambda$ models with parameters given in Table \ref{tableP} and Planck 2015 data \cite{2016A&A...594A..13P}. The bottom panel shows the deviations of the models from the Planck data.}
\label{fitall}
\end{figure}

Table \ref{tableP} shows best fit parameters for both $\Lambda$CDM cosmology and cosmology with Everpresent $\Lambda$. We see that for $\alpha = $0.8824 and $\mu=$ 0.9804 there exists a history of dark energy that is a good fit to the CMB + BAO.  Figure \ref{fitall} plots the anisotropy power spectrum $D_{\ell}^{TT}$ versus $\ell$, for the two models.
%
\begin{table}                   
\centering
\begin{tabular}{|c|c|c|}
 \hline
 Cosmological & & \\
 Parameters & $\Lambda$CDM & Everpresent $\Lambda$  \\
 \hline
 $\Omega_b h^2$ &0.02225 $\pm$ 0.00019 & 0.02205 $\pm$  0.00021  \\
 \hline
 $\Omega_c h^2$ &  0.11857 $\pm$ 0.0012 & 0.1208 $\pm$ 0.0026   \\
  \hline
 100$\theta_{MC}$ & 1.04 $\pm$ 0.00040 & 1.041 $\pm$ 0.00063\\
  \hline
 $\tau$ & 0.06782 $\pm$ 0.012 &  0.06903  $\pm $0.014  \\
 \hline
 $n_s$ &  0.9675 $\pm$ 0.0043 &  0.9593 $\pm$  0.0064  \\
 \hline
 $10^9 A_s$ &2.146 $\pm$ 0.049 & 2.168  $\pm$ 0.066  \\
 \hline
 $\Omega_\Lambda$ & 0.6920 $\pm$ 0.0072 &  0.6781  $\pm$ 0.0081  \\
 \hline
$H_0$ &  67.79$ \pm$ 0.54 & 68.48 $\pm$  0.67 \\
 \hline
 Age/Gyr & 13.80 $\pm$ 0.027  &   13.77 $\pm$  0.031 \\
\hline
$\alpha$ & $-$ & 0.8824  \\
\hline
$\mu$ & $-$ &  0.9804  \\
\hline
$\chi_{\rm model-data}^2 $ & 11334.6 & 11335.2 \\
\hline
\end{tabular}
\caption{Parameters for $\Lambda$CDM cosmology  and cosmology with Everpresent $\Lambda$ (model 2) computed from the 2015 baseline Planck (in combination with lensing reconstruction and BAO (6DF, MGS, DR11CMASS, DR11LOWZ, DR11LyaAuto, DR11LyaCross)) likelihoods, using CosmoMC. This illustrates the consistency of parameters determined from the temperature and polarization spectra at high multipoles. }
\label{tableP}
\end{table}

\begin{table}                   
\centering
 \begin{tabular}{|c|c|c|}
\hline
& & \\
& $\Lambda$CDM & Everpresent $\Lambda$  \\
\hline
CMB: BKPLANCK  & 45.117 & 44.636\\
\hline
CMB: lensing  & 12.157 &13.930 \\
\hline
plik  & 1164.783 & 1165.278\\
\hline
lowTEB   & 10098.575 & 10097.773 \\
\hline
BAO: 6DF  & 0.087 & 0.291 \\
\hline
BAO: MGS  & 0.927 & 2.217 \\
\hline
BAO: DR11CMASS &  2.856 & 3.015  \\
\hline
BAO: DR11LOWZ &  1.098 &1.442 \\
\hline
BAO: DR11LyaAuto  & 4.265 & 2.636 \\
\hline
BAO: DR11LyaCross  & 4.748 & 3.945\\
\hline
& & \\
Total & 11334 & 11335 \\
\hline
\end{tabular}
  \caption{Breakdown of $\chi_{\rm model-data}^2$'s into different datasets, for $\Lambda$CDM and best-fit Everpresent $\Lambda$.}
  \label{tableQ}
\end{table}

\begin{figure}
\centering
\includegraphics[scale=1.4]{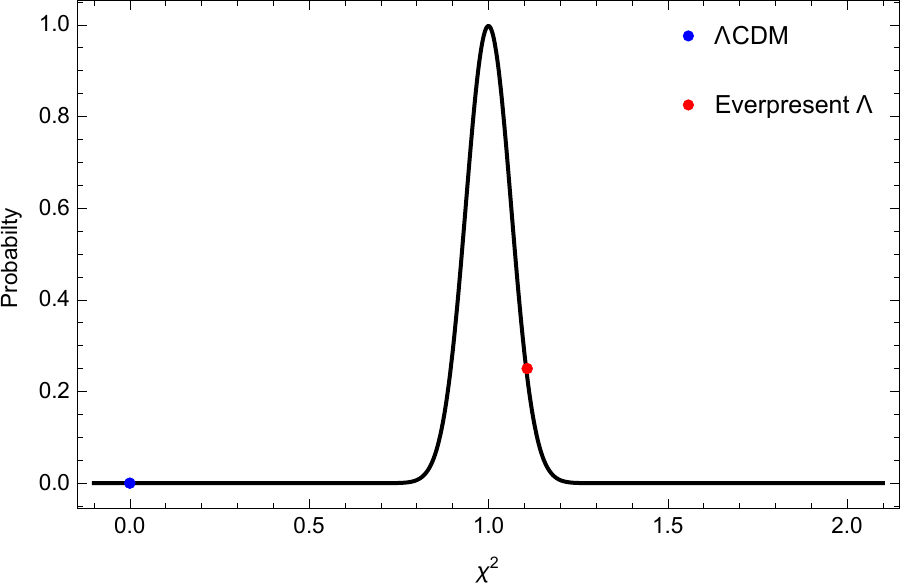}
\caption{$\chi^2_{red}$ for stochastic dark energy with 1000 $\omega$'s sampled. This shows that the best-fit history is a typical model-2 realization, whereas $\Lambda$CDM would be a very atypical outcome of model-2.}  \label{chi2f}
\end{figure}

To check how often one can get such a good fit from Everpresent $\Lambda$, we can examine the probability distribution of $\chi^2_{red}$ for Everpresent $\Lambda$.

\begin{align*}
  P\left[\chi^2_{red}(\tilde{\Omega}_{de})\right]
  \propto
  \int D\tilde{\Omega}'_{de} \, \exp\left(-\frac{\chi^2(\tilde{\Omega}'_{de})}{2}\right)
   \delta_D\left[\chi_{red}^2 (\tilde{\Omega}'_{de})-\chi^2_{red}(\tilde{\Omega}_{de})\right] \ ,
\end{align*}
where $\chi^2_{red}(\tilde{\Omega}_{de})$ is given by Equation \ref{chi2e} (not to be confused with $\chi_{\rm model-data}^2$  in Tables \ref{tableP}-\ref{tableQ} that quantifies how well the model in question fits the data).
Figure \ref{chi2f} shows the expected $\chi^2_{red}$ distribution from gaussian statistics, as well as $\chi^2_{red}(\tilde{\Omega}_{de})$ for $\Lambda$CDM and best-fit  Everpresent $\Lambda$ dark energy history $\tilde{\Omega}_{de}(z)$. ($\Lambda$CDM has a small $\chi^2_{red}$ because most of its $\tilde{\Omega}_{de}$ are zero.) We see that the best fit Everpresent $\Lambda$ realization is only $1.8\sigma$ away from the mean, while $\Lambda$CDM sits well out in the tail at $\sim 4.3\sigma$. This signifies that, judged by their $\chi^2$ statistics, the random gaussian dark energy histories that fit the current data are not atypical. We will have to revise this assessment, however, if future observations prefer histories closer and closer to $\Lambda$CDM.

Fixing the Model 2 parameters at $\alpha=0.8824$ and $\mu=0.9804$, how likely is it to get a universe like ours in Figure \ref{Reco}, amongst all possible realizations? Since the typical value of $|\Omega_{de}|$ is 0.7 for this model, there is nothing unusual about the current density of dark energy. The only peculiar feature is that dark energy happens to be relatively small ($|\Omega_{de}|<0.25$) near last scattering, $z \simeq 1000$, where most the CMB signal originated. However, given the value of $\alpha$, this can happen with probability $\sim 23\%$ which is not particularly unlikely.


\section{Local Measurements of the Hubble Constant} \label{local_hubble}
The past few years have seen local measurements of the Hubble constant diverge from
values inferred from Planck CMB and BAO observations \cite{2017NatAs...1E.169F}. For example, direct measurements of the Hubble constant $H_0$ using Ia supernovae by Riess \textit{et al.} \cite{LocalHub} yielded $H_0 = 73.24\pm 1.74 \text{km s}^{-1} \text{Mpc}^{-1}$~, which is 3.4 $\sigma$ higher than the value of $H_0 = 67.8\pm 0.9 \text{km s}^{-1} \text{Mpc}^{-1}$ yielded by $\Lambda$CDM best fits to the Planck data \cite{2016A&A...594A..13P}.
The Hubble constant is given by
\begin{align}
   \log H_0 = \frac{M_x^0 + 5a_x + 25}{5}
\end{align}
where $M_x^0$ is the absolute magnitude luminosity,
$a_x$ is the intercept of the SN Ia magnitude-redshift-relation $a_x \sim \log cz - 0.2m_x^0$ and $m_x^0$ is the apparent magnitude flux of luminosity\footnote{\textit{Apparent magnitude, m} is a measure of apparent brightness related to the flux $m = -2.5 \log (f/f_0)$ where $f_0$ is the flux that would have apparent magnitude 0.  \textit{Absolute magnitude}  is a measure of intrinsic brightness related to luminosity \cite{Apparent}} . 
Riess \textit{et al.} \cite{LocalHub} argue that this discrepancy in the value of $H_0$ cannot be due to 
inhomogeneities in $\Lambda$CDM cosmology, since simulations that account for inhomogeneities never change $H_0$ by more by 1.3 times statistical uncertainties.

\begin{figure}
  \centering
  \includegraphics[scale=1]{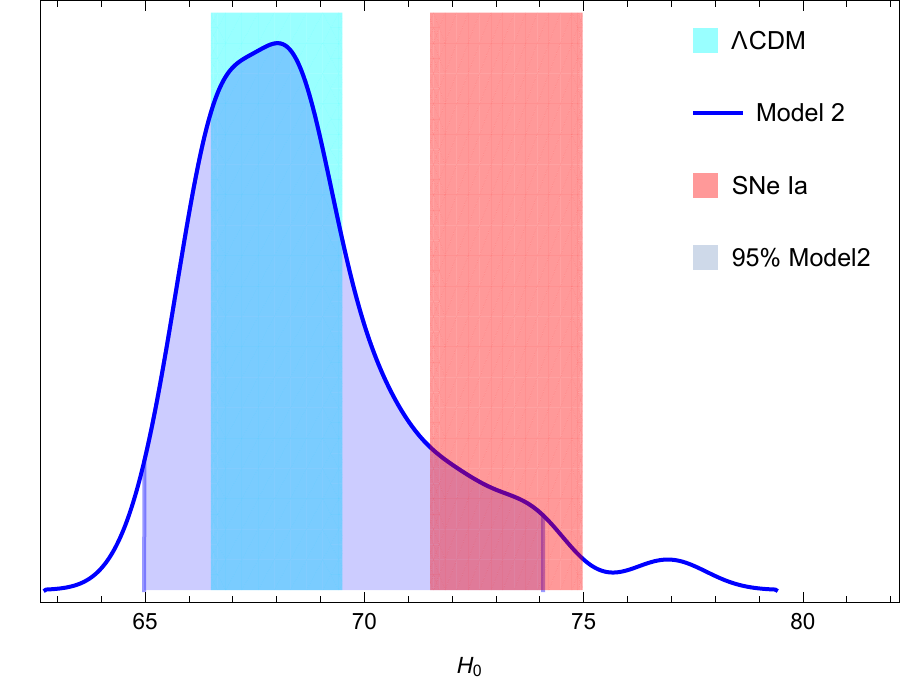}
\caption{Comparison of 2$\sigma$
constraints on $H_0$ from SNe Ia direct measurements from Riess \textit{et al.} \cite{LocalHub} (red region) with those from Planck+BAO data \cite{2016A&A...594A..13P} derived within $\Lambda$CDM (blue region), and within Everpresent $\Lambda$  Model 2 (grey region). We see that the tension between Planck+BAO and SNe Ia is relieved in the Everpresent $\Lambda$ model.}\label{localh0}
\end{figure}

In this regard, Everpresent $\Lambda$ fares better than $\Lambda$CDM, because with a fluctuating dark energy,
local estimates of
$H_0$ are less constrained by high redshift data like CMB or BAO. As seen in Figure \ref{localh0}, higher values of $H_0$ are consistent with the Planck+BAO data; and the SNe Ia measurements fall within 2$\sigma$ of Model 2.

\section{BAO Measurements} \label{sec3}
Baryon acoustic oscillations (BAO) revealed by galaxy surveys probe the expansion history of the universe and provide evidence for dark energy which is independent of that from type Ia supernovae. Before last scattering, baryons were coupled to photons, forming a plasma, and initial perturbations in the density of the baryon-photon plasma propagated acoustically with the speed of sound. Since dark matter was cold, however, the same initial perturbations in dark matter did not propagate acoustically. After last scattering, the baryon
perturbations also stopped propagating, but (over time) they left a gravitational imprint on dark matter, as baryons would have comprised 16\% of the total mass density.
As a result, dark matter (and the resulting galaxy) distributions exhibit a characteristic structure (reflected in correlation functions) on the scale $r_d$ of the sound horizon at last scattering.
%
%
This structure provides a
comoving ruler that is independent of redshift or orientation.

Measuring large scale structure modes perpendicular to the line of sight gives angular diameter distance, $D_{A}$, and measurements along the line of sight determine $D_{H}(z) = \frac{c}{H(z)}$. Combining all modes to suppress noise gives $D_v$,
\begin{align}
   D_v(z) &= \left( z (1+z)^2 D_H(z) D_A(z)^2 \right)^{1/3}   =  \left( z D_H(z) D_c(z)^2 \right)^{1/3},
\end{align}
where $D_c(z)$ is the co-moving distance. Assuming a spatially flat FRW geometry, we have $D_c (z)=\int_0^z \frac{c}{H(z)} dz$~.

It should be borne in mind, however, that most such analyses of BAO data presuppose a fiducial cosmological model which is used to assign flux pairs separated in angle and wavelength to comoving distances, and thereby to fit
the positions of the observed acoustic peaks. The sound horizon $r_d$ is constrained by the locations of these peaks (not their heights), these locations having been determined by pre-recombination physics.

\begin{figure}[h!]
\centering
\includegraphics[scale=1.4]{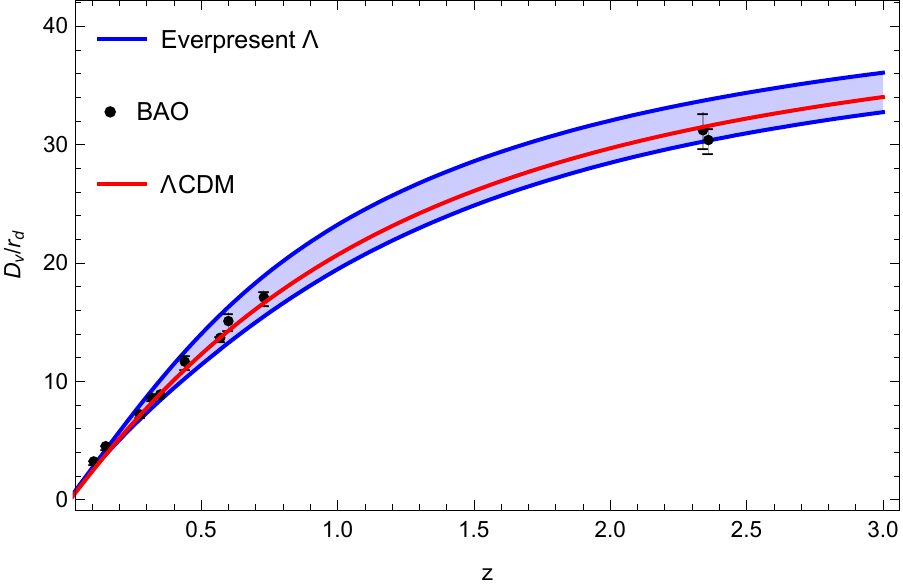}
\caption{BAO measurements from Table \ref{dataT} and model predictions. $\Lambda$CDM is in red, while gray is 68\%  region from Everpresent $\Lambda$ with dark energy histories that give a relatively good fit to the CMB.}
\label{Daplot}
\end{figure}

Unfortunately the fiducial models on which most BAO analyses in the literature are based take dark energy to be negligible before recombination.  In contrast, Everpresent $\Lambda$ models violate this assumption almost by definition, albeit the histories that best fit the data exhibit an $\Omega_{de}$ which is not too far from zero at the time of recombination. Since a fiducial model with non-negligible pre-recombination dark energy could yield different BAO distance measurements, it seems possible that most analyses are introducing an implicit bias against an early dark energy.  Be that as it may, the recent BAO analyses are compatible with Everpresent Lambda or even favour it.


\begin{table}
  \begin{footnotesize}
  \begin{tabular}{|c|c|c||c|c|c|}
  \hline
  & $z$ &Distance(Mpc)& $ $& $z$ &Distance(Mpc)  \\
  \hline
  6dF ($D_v$) \cite{Data1} &0.106 & 457 $\pm$ 27  & SDSS DR9 LRG ($D_A$)\cite{Data6}  &0.57 & 1386 $\pm$ 45\\
  \hline
   SDSS DR7 ($D_v$)  \cite{Data2} &0.15 & 664 $\pm$ 25 & WiggleZ ($D_v$)  \cite{Data7}& 0.6 & 2221 $\pm$ 101 \\
  \hline
   SDSS DR7+2dF ($D_v$) \cite{Data3}  & 0.275 & 1059 $\pm$ 27 & WiggleZ ($D_v$) \cite{Data7} & 0.73 & 2516 $\pm$ 86  \\
  \hline
  SDSS DR11 ($D_v$) \cite{Data4} & 0.32 & 1264 $\pm$ 25  & Ly$\alpha$ auto-corr ($D_A$) \cite{BAO2} & 2.34 & 1662 $\pm$ 96 \\
  \hline
  SDSS DR7 LRG  ($D_v$) \cite{Data5}  &0.35 & 1308 $\pm$ 25 & Ly$\alpha$ auto-corr ($D_H$) \cite{BAO1} & 2.36 &  226 $\pm$ 8 \\
  \hline
  WiggleZ ($D_v$)  \cite{Data7} & 0.44 & 1716 $\pm$ 83  & Ly$\alpha$ auto-corr  ($D_A$) \cite{BAO1} & 2.36 & 1590$\pm$60\\
  \hline
  \end{tabular}
  \caption{BAO data: Volume weighted distance $D_v$ for different $z$ from several sky surveys. } \label{dataT}
  \end{footnotesize}
\end{table}

Figure \ref{Daplot} shows BAO measurements from Table \ref{dataT}, along with Everpresent $\Lambda$ and $\Lambda$CDM predictions (using the cosmological parameters in Table \ref{tableP}).

\begin{figure}
\centering
\includegraphics[scale=0.4]{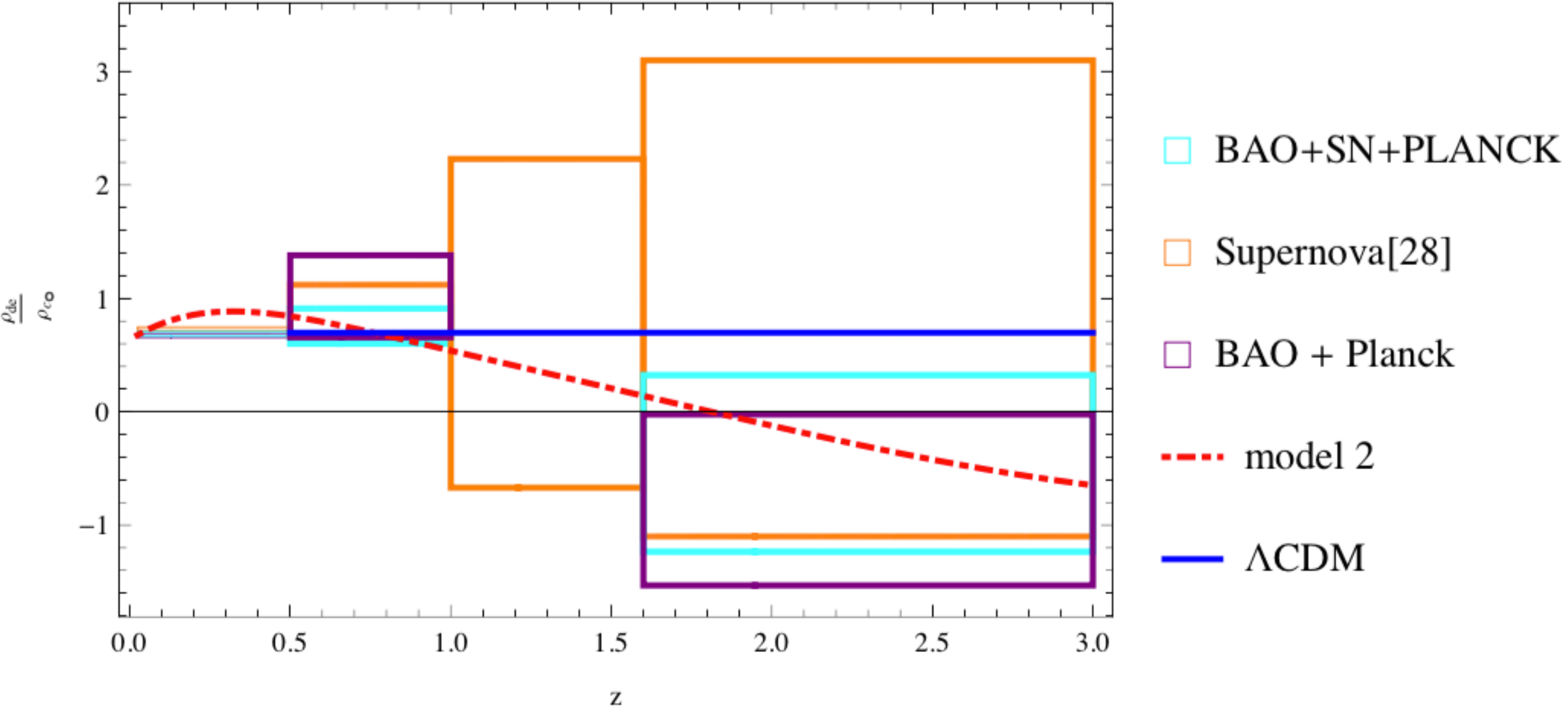}
\caption{Model histories of $\Omega_{de}$ compared with BOSS 68\% bounds from \cite{BAO3}. The vertical axis is $\rho_{de}$ in units of its value now.}  \label{bossplot}
\end{figure}

Figure \ref{bossplot} shows that the dark energy history in Everpresent $\Lambda$ that best fits the Planck 2015 CMB data \cite{2016A&A...594A..13P} falls well within the BOSS bounds \citep{BAO3}, which on the other hand are in $2.5\sigma$ tension with $\Lambda$CDM.
Indeed, if we fix the cosmological parameters to their best fit values in Table \ref{tableP},
there is a 78$\%$ chance that $\Omega_{de}$ is negative
somewhere in
the range $1.5 < z < 3$. This is consistent with the negative $\Omega_{de}$ found by Aubourg \textit{et al.} \cite{BAO3} and seen in Figure \ref{bossplot}.


\section{Ultramassive Black Holes at High Redshifts }\label{sec5}
There is overwhelming observational evidence that the centers of all large galaxies host supermassive black holes which grow by accreting the surrounding gas. The most efficient phase of this growth leads to bright active galactic nuclei, known as quasars, with luminosities close to the astrophysical upper limit for Eddington accretion. However, quasars at $z\sim$ 6 to 7 appear to host black holes more massive than would be allowed by Eddington accretion in a $\Lambda$CDM cosmology, starting from seeds of stellar mass $\sim$ 5 to 20 M$_{\odot}$, \textit{\/assuming that\/} these black holes are maximally rotating. Exotic astrophysical processes might possibly induce super-Eddington accretion rates, allowing the quasars to accumulate mass faster, or they might allow for direct collapse to much more massive black hole seeds.

However, it is also possible (e.g. \cite{Tpro}) that the solution to the puzzle could come from replacing a $\Lambda$CDM cosmology by one that would allow more time for the accretion to take place. Indeed, Everpresent $\Lambda$ produces a range of expansion histories, many of which do give the quasars more time to accumulate mass.

The mass of a blackhole forming by accretion is given by
\begin{align}
M(t) = M_{\text{seed}} \exp\left( \frac{t}{\eta \tau} \right),
\end{align}
where $M_{\text{seed}}$ is the mass of the seed (assumed to be $20M_{\odot}$), $\eta$ is the accretion efficiency, and $\tau = \frac{Mc^2}{L_{Edd}}$ is given by
\begin{align}
\tau = \frac{\sigma_T c X}{4\pi G m_p}  \ ,
\end{align}
with $\sigma_T$ being Thomson scattering cross-section, $c$ the speed of light, $G$ the gravitational constant, $m_p$ the proton mass, and where $X = (1 + 0.75)/2$ if one assumes a Helium mass fraction of 25\%.
The accretion efficiency depends on the spin of the blackhole according to \cite{1973blho.conf..343N}:
\begin{align}
  \epsilon = \frac{\eta}{1-\eta} &= \frac{1}{\epsilon_g}\left( 1 - \sqrt{1 - \frac{2 G M}{c^2 r_{\text{{\scriptsize ISCO}}}}}  \right) \\
  r_{{\scriptsize ISCO}} &= \frac{GM}{c^2}\left( 3 + \alpha_2 - \sqrt{(3-\alpha_1)(3+\alpha_1+2\alpha_2)} \right) \\
  \alpha_1 &= 1 + (1-a_*^2)^{1/3} \left( (1+a_*)^{1/3} + (1-a_*)^{1/3}  \right) \\
  \alpha_2 &= \sqrt{3a_*^2 + \alpha_1^2}
\end{align}
where $a_*\in[0,1]$ is the dimensionless spin parameter.

\begin{figure}[h!]
\centering
\includegraphics[scale=0.365]{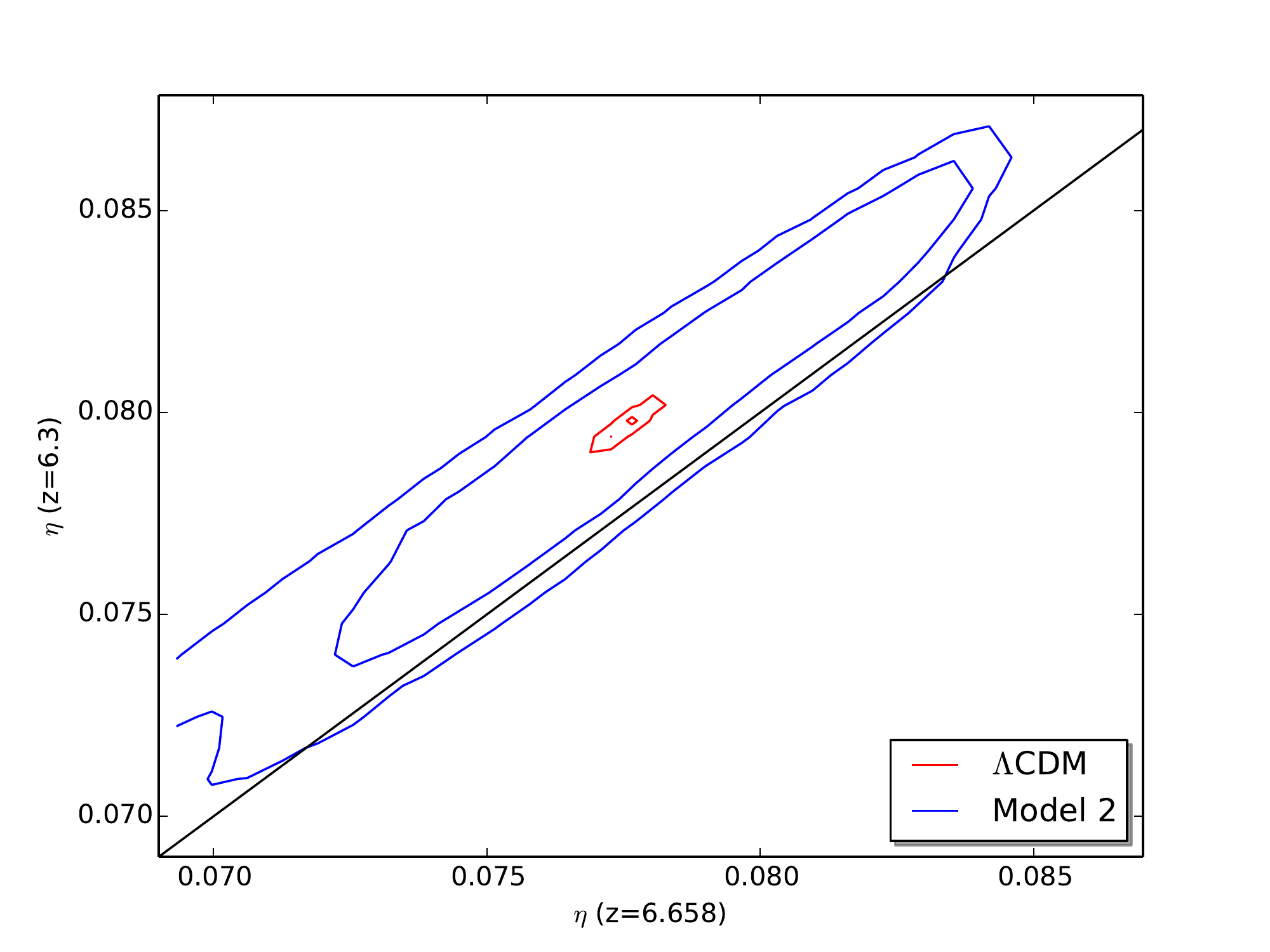}
\includegraphics[scale=0.365]{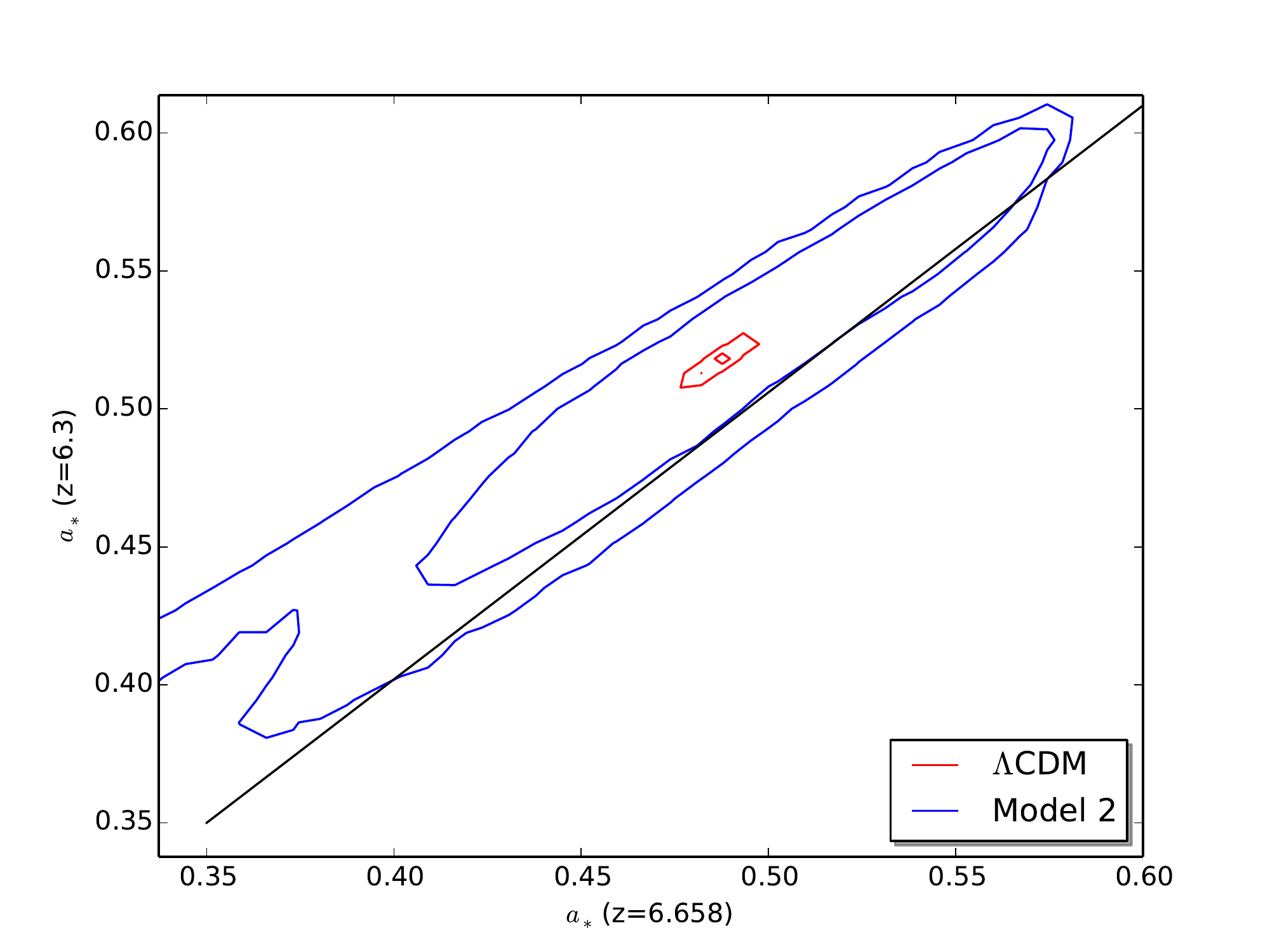}
\caption{68\% and 95\% constraints on accretion efficiency $\eta$ and on spin-parameter $a_*$ for a pair of quasar-blackholes, shown for $\Lambda$CDM and Everpresent $\Lambda$ (Model 2). For neither model can the blackholes be maximally spinning ($a_*=1)$, although Everpresent $\Lambda$ allows for wider ranges of spins and efficiencies.}
\label{massk}
\end{figure}

Figure \ref{massk} shows bounds on the accretion efficiencies and on the spin-parameters
for two particular black holes at redshifts  6.30 and 6.658 respectively,
namely those of the quasars J010013.02+280225.8 \cite{BHmass1} and J338.2298+29.5089 \cite{Tpro6}. In neither $\Lambda$CDM nor Everpresent $\Lambda$ can the black holes be maximally spinning ($a_* = 1$), but Everpresent $\Lambda$ does allow for higher spins and efficiencies.

\section{Primordial Lithium-7 Problem} \label{sec6}
Models of Big Bang Nucleosynthesis (BBN) predict the abundance of light cosmological elements (D, $^{3}$He, $^{4}$He, $ ^{7}$Li) that are produced in the first $10$ seconds to $20$ minutes after the Big Bang, when the universe was still hot and dense. The astronomically observed abundances of Deuterium and Helium-4 are in agreement with BBN predictions, while the predicted Helium-3 abundance is also within observational bounds. However, three times more Lithium-7 is predicted than observed, this being known as the {\it\/Lithium problem\/}.

Fields \cite{Li7pro} discusses a number of possible solutions to this problem. One solution might be that the Lithium-7 present in halo stars does not reflect the
initial abundance because
some of the Lithium-7 was destroyed through nuclear binding. For some stars Lithium destruction has been studied, and some groups \cite{Li7distroy} have found a certain amount of Lithium depletion. Though this might be a direction toward resolving the Lithium problem, it is far from clear, since in metal-poor stars (where less depletion is expected) Lithium-7 abundance is nowhere near the predicted value. 

The problem might be in the nuclear physics, if BBN calculations have been leaving out some reactions or using incorrect rates. But few reactions have been found to be relevant for producing light elements, and this has been studied in the laboratory. Another solution might be that dark matter introduces new processes which can alter light element abundances during or after BBN \cite{Pospelov:2010hj}.

Could Everpresent $\Lambda$ solve the problem by changing cosmological expansion rates? The expansion rate is determined by the Friedmann equation,
\begin{align}
   H^2 &= \frac{8\pi G}{3} \left(\rho + \Lambda(t)\right) \ .
\end{align}
We ran Timmes's BBN code \cite{BBNcode} using values of $N_{\text{eff}}$ and  $\eta$ from Planck 2015. Here, $N_{\text{eff}}$ is the effective number of degrees of neutrino species during BBN,
and $\eta$ is the photon-to-baryon ratio,\footnote{Lower-case $h$ is the dimensionless Hubble parameter, $h = $ Hubble Parameter/($100$km s$^{-1}$Mpc$^{-1}$).}

\begin{align}
   \eta_b &=2.7377\times 10^{-8} \Omega_b h^2  \ ,
\end{align}

which depends on $\Omega_b h^2$. For $\Lambda$CDM we took $N_{\text{eff}}=3$ and $\Omega_b h^2$ from Table \ref{tableP}. For Everpresent $\Lambda$ we took $N_{\text{eff}}=3$ and sampled $\Omega_b h^2$ from Markov chains
that had relatively good $\chi^2$ relative to the CMB data.\footnote%
{Although we simply fixed $N_{\text{eff}}$ to 3, it would be more correct to fit for it.}
The results are displayed in Table~\ref{tableBBN}

\begin{table}
   \centering
   \begin{tabular}{|c|c|c|c|c|c|}
   \hline
   & Observation  & Everpresent $\Lambda$ &  $\Lambda$CDM \\
   \hline
   D/H ($\times 10^{-5}$) & 2.547 $\pm 0.033$ & 2.64  &  2.584$_{-0.035} ^{+0.036}$ \\
    \hline
   $^3$He/H ($\times 10^{-5}$)& 1.1 $\pm 0.2 $& 1.09  & 1.026$_{-0.006} ^{+0.005}$\\
   \hline
   $^4$He/H & 0.249 $\pm 0.009$ & 0.2311   & 0.248 $\pm$ 0.001 \\
    \hline
   $^7$Li/H ($\times 10^{-10}$) & 1.23$^{+0.34}_{-0.16}$   & 3.583   & 4.507$\pm0.08$  \\
   \hline
   $\chi^2$ & & 57  & 89  \\
   \hline
   \end{tabular}
   \caption{Predictions for primordial cosmological abundances of light elements from BBN with different cosmologies. The uncertainties in the case of $\Lambda$CDM come from uncertainties in the baryon density.}
\label{tableBBN}
\end{table}

Deuterium is measured by observing the absorption of hydrogen in quasar spectra. Several groups have measured D/H abundance from different quasars at different redshifts. The highest bound for $D/H$ is given by Burles and Tytler \cite{Do4} D/H =3.39$\pm 0.3\times 10^{-5}$ and the lowest by Kirkman \textit{et al.} \cite{Do2} $\text{D/H}=2.42^{+0.35}_{-0.25} \times 10^{-5}$, measured from light from quasar 1009+2956 at $z$=2.504. In Table \ref{tableBBN} the measurements with the lowest uncertainties are used, Cooke \textit{et al.} \cite{DH2016}. Bania \textit{et al.} \cite{He3o1} determined primordial Helium-3 abundance in the Milky way, setting the highest bound to be $^3\text{He/H} = 1.1\pm 0.2 \times 10^{-5}$. Astronomical observations of metal poor halo stars\cite{Li7o1,Li7o3,Li7o4} give relative primordial abundance of $^7\text{Li/H} = 1.23^{+0.34}_{-0.16} \times 10^{-10}.$ Olive \textit{et al.} \cite{He401} determined primordial Helium-4 abundance to be $^4\text{He/H} = 0.249\pm 0.009.$

For Everpresent $\Lambda$ with a given value of $\eta_b$, different dark-energy histories predict different elemental abundances. The histories that correctly predict Lithium-7 predict an abundance of Deuterium  exceeding the highest observational bound. While the best-fit Everpresent $\Lambda$ model is thus not very good, it still does better than $\Lambda$CDM. This suggests that more conservative values for light element abundance measurement-errors might eventually yield a satisfactory fit.
%


\begin{figure}
\centering
\includegraphics[scale=1.3]{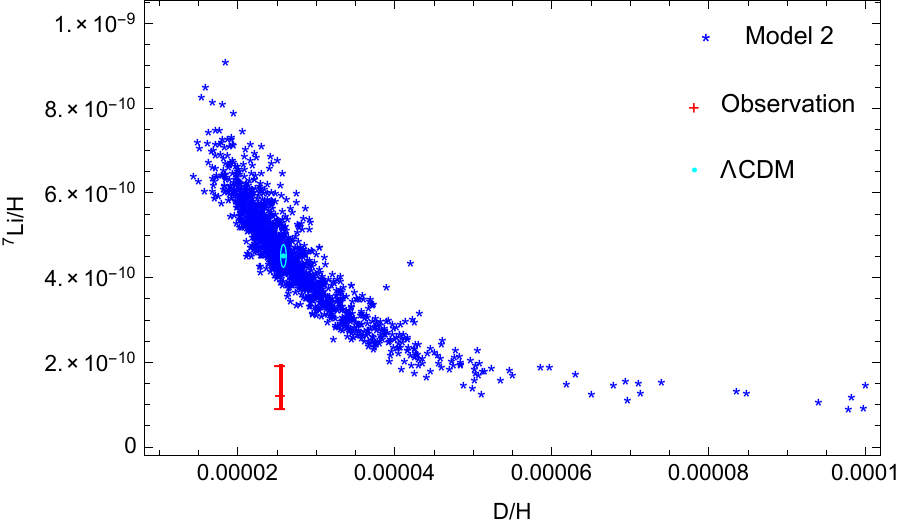}
\caption{Deuterium vs Lithium-7 abundance as predicted by Everpresent $\Lambda$ and $\Lambda CDM$. For Everpresent $\Lambda$, when less Lithium-7 is produced there is more Deuterium.}
 \label{BigbangN}
\end{figure}

\begin{figure}
\centering
\includegraphics[scale=1.3]{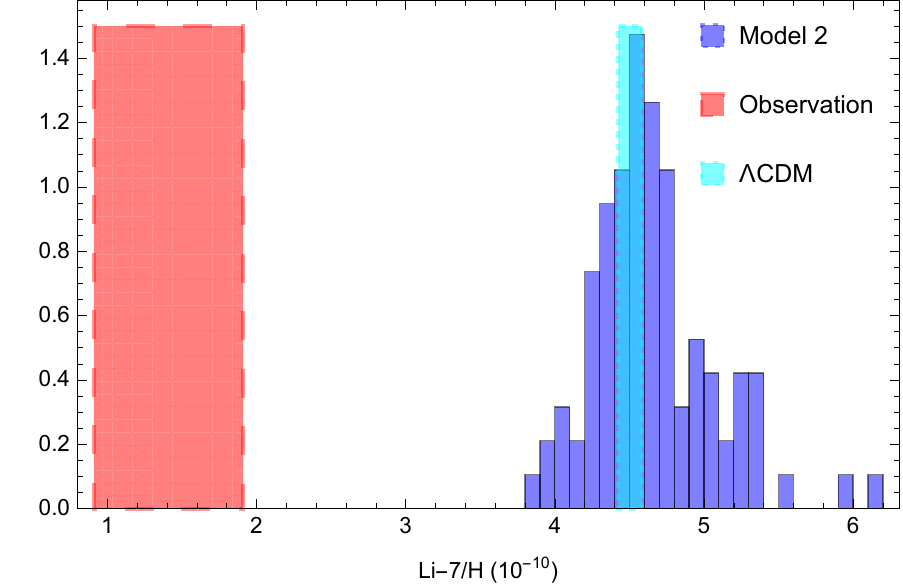}
\caption{Model 2 dark-energy histories that match the Deuterium abundance have a higher Lithium-7 abundance than observed. The blue region is the probability distribution for Lithium-7 in model 2, conditioned on a Deuterium abundance of $2.547\pm0.033$. The red region is the observed Lithium-7 abundance, and the cyan region is the Lithium-7 abundance from $\Lambda$CDM.}
\label{BigbangHist}
\end{figure}

%
\section{Conclusion and Future Prospects}\label{sec_conclude}
The hypotheses that underpin causal set theory led to the prediction that the cosmological ``constant'' would in fact not be constant, but would fluctuate with a magnitude comparable to that of the critical density at any given cosmological epoch.  This implied in particular that the contemporary value of $\Lambda$ would be found to be comparable to the Hubble scale, a forecast that has in a sense been confirmed by the various pieces of evidence for the so called dark energy.  But of course this is no more than a first step to a fuller theory.

To go further, one needs (since a model based on first principles is not yet available) at least a phenomenological model of fluctuating $\Lambda$ which can be compared with observations.  In this paper we have described two such models, the first being older and closer to the original idea, the second being slightly more general and somewhat easier to simulate.  Most of the results reported above were obtained from the second model.

By definition, any model of a fluctuating $\Lambda$ is inherently stochastic.  As such, it will necessarily have many realizations that differ drastically from our own cosmos.  However, the simulations discussed above show that there are also many realizations that broadly resemble reality, including some that agree as well as (or better than) $\Lambda$CDM does with several pieces of observational evidence, including CMB, BAO, and SNe Ia measurements, primordial nuclear abundances, and the very early occurrence of ultramassive black holes. Indeed as we saw, most notably in the discussions of BAO and local measurements of the Hubble constant, some of the ``tensions'' that have arisen between $\Lambda$CDM and observations can be removed in an Everpresent $\Lambda$ cosmology.

For the CMB, there is a good fit to Planck 2015 data by a realization coming from the Model-2 parameters, $\alpha = 0.8824$ and $\mu=0.9804$. Of course (and as just emphasized), this does not mean that every history of dark energy with $\alpha = 0.8824$ and $\mu=0.9804$ will fit the CMB, nor does it imply that only histories with $\alpha = $0.8824 and $\mu=$ 0.9804 can do so.

If $\Lambda$ really is fluctuating and ``everpresent'', that should become apparent as observations accumulate.  Meanwhile it seems important to understand to what extent the good results we have obtained for the CMB depend on the details of how we have handled the spatial inhomogeneities that the variations in the CMB reflect.  To the overall picture of a $\Lambda$ that varies temporally with a coherence time set by the Hubble scale, we have added some ad hoc assumptions made solely for the sake of achieving a mathematically consistent perturbation scheme.  In particular, we have relied on a $\mathtt{CAMB}$ code that effectively models dark energy as a scalar field (``quintessence''), which is not fully in accord with the original idea according to which the fluctuations in $\Lambda$ reflect an underlying dynamics that is both nonlocal and quantal in character.

We would hope that our results would be relatively insensitive to our ad hoc assumptions.  We would even expect this because of the separation of scales between the fluctuations in $\Lambda$ and the fluctuations that are relevant to the CMB.  But to test how robust our results really are, it would be well to repeat our simulations with similar models that realize the same basic idea, starting perhaps with a more quantitative comparison between Model 2 and Model 1 (which we have similated only enough to confirm that it behaves qualitatively like Model 2.)   One could also, for example, ``turn off the dark-energy perturbations'' in $\mathtt{CAMB}$).

But ultimately one should seek to incorporate spatial inhomogeneities in a less ad hoc manner which does justice to the nonlocal character of the underlying fluctuations, and which therefore does not resort to modelling $\Lambda$ as a fluid (and specifically a fluid with a realization-dependent equation of state.)

\section{Acknowledgments}
The authors are grateful to Alireza Hojjati, Valeria Pettorino and Elizabeth Gould for helpful discussion on using CosmoMC code, as well as Anze Slosar and Jose Vazquez for helpful discussions on BAO measurements. The authors would also like to thank Fay Dowker and Maqbool Ahmed for helpful discussions on Everpresent $\Lambda$, and an anonymous referee for suggesting a number of improvements in presentation.

This research was supported in part by NSERC through grant RGPIN-418709-2012. This research was supported in part by Perimeter Institute for Theoretical Physics. Research at Perimeter Institute is supported by the Government of Canada through the Department of Innovation, Science and Economic Development Canada and by the Province of Ontario through the Ministry of Research, Innovation and Science.

\bibliography{fcc_3}{}
\bibliographystyle{JHEP}

\end{document}